# Highly efficient source for frequency-entangled photon pairs generated in a 3$^{rd}$ order periodically poled MgO-doped stoichiometric LiTaO$_3$ crystal

Heonoh Kim, Hee Jung Lee, Sang Min Lee, and Han Seb Moon*

*Department of Physics, Pusan National University, Geumjeong-Gu, Busan 609-735, Korea*
*Corresponding author: hsmoon@pusan.ac.kr*

We present a highly efficient source for discrete frequency-entangled photon pairs based on spontaneous parametric down-conversion using 3$^{rd}$ order type-0 quasi-phase matching in a periodically poled MgO-doped stoichiometric LiTaO$_3$ crystal pumped by a 355.66 nm laser. Correlated two-photon states were generated with automatic conservation of energy and momentum in two given spatial modes. These states have a wide spectral range, even under small variations in crystal temperature, which consequently results in higher discreteness. Frequency entanglement was confirmed by measuring two-photon quantum interference fringes without any spectral filtering.

Entangled photon pairs are a very important resource in the field of quantum information science. Over the last three decades, various types of entangled photon-pair sources have been developed not only to study the foundation of quantum mechanics but also to implement real world quantum technologies [1,2]. Among them, polarization-entangled pairs of photons have been adopted as the most versatile light source because their polarization has a well-defined orthogonal basis, and which can be easily manipulated using only linear optics [3,4]. In combination with the polarization degree of freedom, the frequency (or wavelength) of a single photon can also be employed as a basic unit to prepare a quantum state [5-10]. Particularly, generation and manipulation of the multi-dimensional quantum states by utilizing frequency degree of freedom is one of the interesting issues for potential applications in practical quantum communication system [11-14]. To utilize the frequency degree of freedom as a quantum information resource, it is necessary to separate the center frequencies of the two photons over a range substantially larger than the spectral bandwidth of the two individual photons.

Recently, certain types of photonic sources for discrete frequency-entangled states using correlated photon pairs were reported and used to explore the potential applications in quantum information and communication technologies. In the first experimental demonstration, Ramelow and co-workers reported a discrete tunable frequency-entangled photon-pair source generated by spontaneous parametric down-conversion (SPDC) using type-II collinear quasi-phase matching (QPM) in a periodically poled KTiOPO$_4$ (PPKTP) crystal inside a polarization Sagnac interferometer [5,6]. To successfully obtain frequency symmetric states in the two spatial modes, they used a method of transferring the polarization entanglement onto a frequency-entangled state. Further work was performed using optical fibers, such as a dispersion shifted fiber [7,8] and a commercial polarization-maintaining fiber [9], to generate frequency-entangled photon-pair states via spontaneous four-wave mixing (SFWM). Occasionally, fiber sources have inevitably required cumbersome cooling systems to reduce any accompanying redundant background photons, and also require a filter based frequency post-selection. In the field of integrated quantum photonics, an on-chip frequency entangled photon-pair source was generated through SFWM inside compact silicon waveguides to implement on-chip and off-chip two-photon quantum interferences [15].

In this paper, we present a highly efficient frequency-entangled photon-pair source based on SPDC using a 3$^{rd}$ order type-0 QPM in a periodically poled MgO-doped stoichiometric LiTaO$_3$ (PPMgSLT) crystal. The PPMgSLT crystal employed in our experiments has certain advantages for generating correlated photons. For example, the higher nonlinear coefficient ($d_{33}$=13.8 pm/V) compared with other nonlinear materials requires a relatively lower pump power to generate abundant photon pairs. Moreover, it is possible to produce the correlated photons in the visible spectral range due to the lower ultra violet absorption of the crystal. In our experiment, we used a 20-mm-long PPMgSLT crystal with a poling period of 6.07 μm to obtain a noncollinear type-0 phase matching condition at a given crystal temperature. Although the effective nonlinear coefficient of the crystal using the 1$^{st}$ order QPM is larger than that of the 3$^{rd}$ order, we employed the 3$^{rd}$ order QPM process because of the fabrication difficulties to achieve a high poling quality for small periods below 3 μm. As a pump source, we used a cw mode single-frequency (<1 MHz) ultra-violet diode pumped solid state (DPSS) laser (Cobolt Zouk™ 355 nm) with a center wavelength of 355.66 nm. The crystal temperature was controlled by a thermoelectric cooler (TEC) with an accuracy of 0.01°C. Temperature control was required because noncollinear quasi-phase matching conditions are obtained near room temperature.

Figure 1 shows a conceptual scheme for generation of frequency-entangled photon pairs with automatic energy and momentum conservation conditions between the two photons in two given spatial modes. A quantum state involving a frequency anti-correlated photon-pair is described in the form,

$$|\Psi\rangle = \iint d\omega_1 d\omega_2 \Phi(\omega_1,\omega_2)(|\omega_1\rangle_{P1}|\omega_2\rangle_{P2} + |\omega_2\rangle_{P1}|\omega_1\rangle_{P2}), \quad (1)$$

where $\Phi(\omega_1,\omega_2)$ is the joint spectral amplitude function, $\omega_1$ and $\omega_2$ represent angular frequencies of individual single photons for the two spatial modes of the photon 1 (P1) and photon 2 (P2), respectively. The two-photon state of Eq. (1) is a typical superposition state of the two spectral modes traveling through the two spatial modes, and is a symmetric state containing two distinguishable spectral components $\omega_1$ and $\omega_2$. In particular, two spectral components have different refractive indices and emerged with different time delays from the crystal. Nevertheless, the two photons are in a symmetric state which results in indistinguishable two-photon amplitudes. This condition enables one to construct an entangled state of Eq. (1). This feature is similar to a scheme that uses polarization-entangled two-photon states [16]. When the two photons with different frequencies from the crystal are combined at the broadband nonpolarizing beamsplitter (BS) as shown in Fig. 1, the coincidence detection probability of the two incoming photons with different frequencies can be expressed as,

$$P(\omega_1,\omega_2) = N\left[1 - Vf(\delta t)\cos(\Delta\omega \delta t)\right], \quad (2)$$

where $N$ is a constant, $V$ is the two-photon fringe visibility, $f(\delta t)$ is an envelope function corresponding to the spectral properties of the detected photons, and the cosine term is related to the beat frequency $\Delta\omega$ between the two photons.

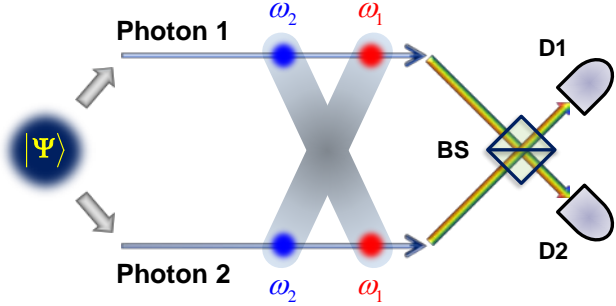

Fig. 1. (a) Generation of frequency-entangled photon pairs in a type-0 PPMgSLT crystal under nondegenerate noncollinear quasi-phase matching conditions with automatic conservation of energy and momentum between two photons in two given spatial modes. Gray bars represent correlated photon pairs with different angular frequencies. Two-photon interference fringes can be observed without any frequency post-selection after the beam splitter (BS).

Full characterization of the photon-pair source can be performed by simply measuring the single and coincidence counting rates as a function of pump power under degenerate and noncollinear phase matching conditions [17]. The degenerate wavelength was 711.32 nm at a crystal temperature of 22.90°C, which was measured using a single mode fiber coupled tunable filter. When the coincidence time window was set at 19.45 ns, the measured coincidence counting rate per input pump power was found to be about 98,500 Hz/mW, which corresponds to about 3 MHz/mW in the photon-pair production rate within the crystal. From the measured result, the conversion efficiency was estimated to be $1.66\times10^{-9}$. The ratio of coincidence to the single counting rate $N_c/\sqrt{N_1 N_2}$ was found to be 18.60±0.30%. The coincidence to accidental coincidence ratio was 16.85×P$^{-1}$, where $N_c$ and $N_{1,2}$ denote the coincidence and single counting rates, respectively, and P is the pump power in mW. The second-order coherence function was measured to be 0.087±0.001/mW. With this source, we have generated frequency-entangled photon-pair sources under a noncollinear phase matching configuration by only controlling the crystal temperature.

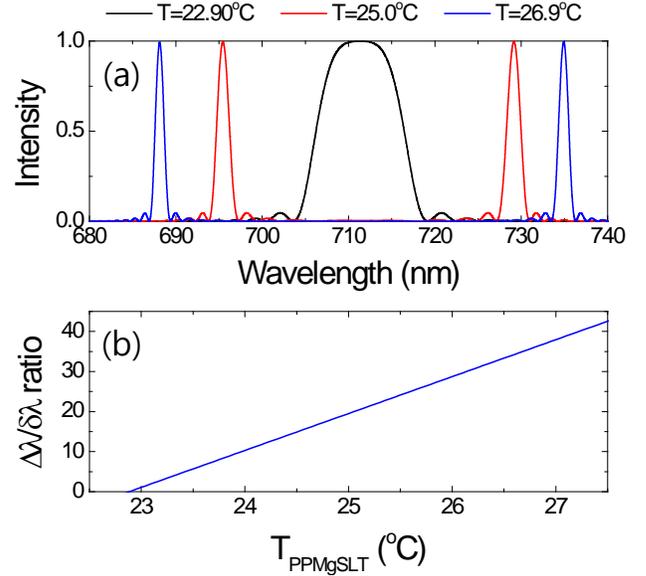

Fig. 2. (a) Theoretically calculated single photon spectra at three different PPMgSLT crystal temperatures. (b) Ratio of the center wavelength difference between two photons (Δλ) to the bandwidth of single photons (δλ), which is obtained from the noncollinear emission spectra of (a).

Successful discrete frequency entanglement can be achieved only when the center wavelengths of the two photons are separated by much more than the spectral bandwidth of the two individual single photons. We have calculated the noncollinear emission spectra of the correlated photons as a function of the crystal temperature, as shown in Fig. 2(a). From this result, we have extracted the ratio of the center wavelength difference between the two photons to the bandwidth of the individual single photons, (Δλ/δλ)/T$_{PPMgSLT}$=9.22/°C, as shown in Fig. 2(b). Compared with type-II PPKTP in Ref. [5,6], (Δλ/δλ)/T$_{PPKTP}$=0.43/°C, our source has a factor of 21 higher than comparable previous works. The PPMgSLT crystal source operates over a wider spectral range under smaller variations in crystal temperature.

To confirm the discrete frequency entanglement of the generated photon pairs, two-photon quantum interference fringes were observed in a Hong-Ou-Mandel (HOM) interferometer [18]. Figure 3 shows the experimental setup. The emission angle of an individual single photon is about 1.7° to the pump beam propagation direction. Pump power was adjusted using a variable neutral density filter (VNDF) before the SPDC crystal. Two photons are collimated by the spherical lens L2 (f=300 mm) and then coupled to the single mode fiber (SMF) using an aspheric lens with a focal length of 8 mm. Two SMFs were connected to the single photon counting modules (SPCM, Perkin Elmer AQR4C). Two spatially separated photons travel through two different paths, and one of the photons is sent to the variable path to adjust the path-length by the prism mounted on the motorized translation state. As is well known, two photons in a symmetric state result in two-photon quantum interference when they overlap at the nonpolarizing beamsplitter (BS), even though they have different properties such as angular frequencies. Furthermore, we emphasized that our two-photon interference experiment is performed without any spectral filtering because our source can satisfy energy and momentum conservation conditions over a wider spectral range in two given spatial modes.

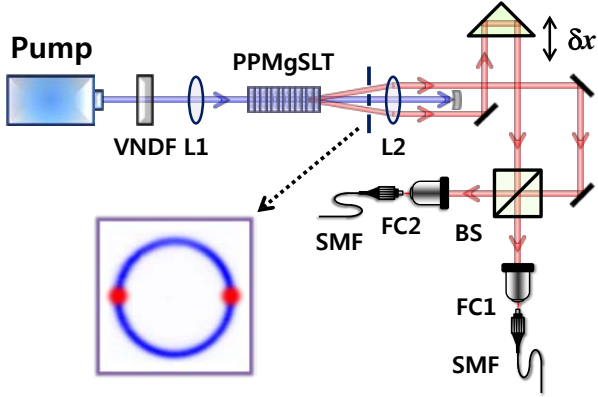

Fig. 3. Experimental setup to observe two-photon interference with frequency-entangled photon pairs. Pump, cw DPSS with wavelength of 355.66 nm; VNDF, variable neutral density filter; L1, L2, spherical lens; BS, nonpolarizing beamsplitter; FC, single mode fiber coupling kit; SMF single mode fiber (SM 600); δx, fine adjustment of the path-length differences. SMF is connected to a single photon detector.

Figure 4 shows the experimental results of the two-photon spatial quantum beating interference [19]. Accidental coincidences were subtracted from the measured coincidence counting rates by using the equation, $N_{acc}=N_1 N_2 T_R$, where $N_1$ and $N_2$ are the single counting rates, $T_R$ is the coincidence resolving time. Circles are normalized coincidence counts as a function of path-length difference between two arms of the HOM interferometer. The solid lines indicate a fit of Eq. (2) to the data points. Eq. (2) yields different spectral properties $f(\delta t)$ of individual photons and for the beat frequencies $\Delta\omega$ between the two single photons. In the experiment, the pump power was set to 23 μW. The degenerate condition was achieved at a crystal temperature of 22.90°C. In this condition, the measured average single and coincidence counting rates were about 24,000 Hz and 1,500 Hz, respectively, within a 19.45 ns resolving time. In this case the ratio of accidental coincidence to the measured coincidence was about 0.72%, which implies that the accidental coincidence had a negligible effect on two-photon interference quality. Visibilities obtained from the data were 90.173±0.003% (22.90°C), 84.031±0.005% (25.00°C), and 84.211±0.006% (26.90°C). In general, two-photon interference visibility was strongly depended on the balance of the transmittance (T) and reflectance (R) of the BS. By measuring T and R of the BS directly (T=60.94±0.06%, R=39.06±0.06%), the maximally attainable fringe visibilities in our experiments was 90.87±0.07%.

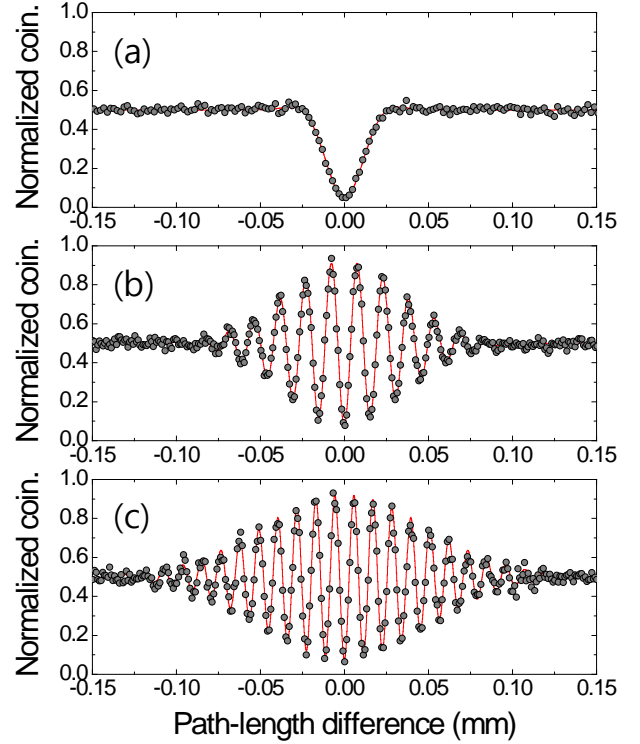

Fig. 4. Normalized net coincidence counts as a function of the path-length difference. Two-photon interference fringes are obtained for three different temperatures of the PPMgSLT crystal (a) T=22.90°C, (b) T=25.00°C, and T=26.90°C.

We now analyze the discreteness of the frequency entanglement by the ratio of the center wavelength difference between two photons to the bandwidth of single photons. In fact, the spectral properties of a single photon can be extracted from the HOM fringes width because we used single mode cw pumped SPDC process [20,21]. From the fitting parameter in Eq. (2), Δλ (δλ) was measured to be 33.01 nm (6.16 nm) at 25.00°C, and 44.72 nm (4.39 nm) at 26.90°C, which results in about $(\Delta\lambda/\delta\lambda)/T_{PPMgSLT}$=2.55/°C. Compared with the PPKTP source [5], the experimentally obtained discreteness improvement was found to be about 6 times, because the estimated value of δλ is restricted by finite acceptance angle of the coupling optics for the single mode fiber used in our experiment. The other cause of the discrepancy between the calculated values in Fig. 2 and the experimental results in Fig. 4 may be due to the inaccuracy of the temperature dependence of the refractive index of the crystal. Furthermore, this not only be caused by an inaccurate absolute value of the TEC reading but also the broader coupling bandwidth of the noncollinear spectra to the SMF. The discrepancy might also arise from the difference in the estimation of the beat frequency $\Delta\omega$ between two photons in Eq. (2).

In conclusion, we have demonstrated for the first time a highly efficient discrete frequency-entangled photon-pair source using a 3$^{rd}$ order type-0 QPM SPDC in a PPMgSLT crystal. Correlated two-photon states with high brightness and high purity were generated in the visible spectral range under noncollinear QPM SPDC conditions, which have a wider spectral range and automatically satisfy energy and momentum conservation in two given spatial modes. Discrete frequency entanglement was confirmed by measuring HOM two-photon interference fringes involving two different wavelengths, which was obtained without any spectral filtering. Frequency-entangled quantum states can be used in quantum information technologies utilizing higher dimensional encoding in the frequency domain, and can also be used in single photon level quantum metrology.

This work was supported by the Basic Science Research Program through the National Research Foundation of Korea funded by the Ministry of Education, Science and Technology (Grant No. 2015R1A2A1A05001819, No. 2014R1A1A2055488, and No. 2014R1A1A2054719). Also, this work was supported by the Measurement Research Center (MRC) Program for Korea Research Institute of Standards and Science. H. Kim and H. J. Lee contributed equally to this work.